\documentclass[12pt]{article}
\usepackage{graphicx}
\input epsf.tex
\def\DESepsf(#1 width #2){\epsfxsize=#2 \epsfbox{#1}}
\def\NPB{{ Nucl. Phys.} B}

\def\PLB{{ Phys. Lett.}  B}
\def\PRL{ Phys. Rev. Lett.}
\def\PRD{{ Phys. Rev.} D}

\def\ZPC{{ Z. Phys.} C}
\def\AP{ Astropart. Phys.}

\def\IJMPA{{ Int. J. Mod. Phys.} A}

\newsavebox{\sboxpubnumber}
\newsavebox{\sboxpubdate}
\newcommand{\pubdate}[1]{\begin{lrbox}{\sboxpubdate}{#1}\end{lrbox}}
\newcommand{\pubnumber}[1]{\begin{lrbox}{\sboxpubnumber}{\begin{tabular}{l} #1 \\
				 \usebox{\sboxpubdate}
				 \end{tabular}}
                           \end{lrbox}
                           \pubblock}
\newcommand{\Title}[1]{\begin{center} {\Large #1 } \end{center}}
\newcommand{\Author}[1]{\begin{center}{ \sc #1} \end{center}}
\newcommand{\Address}[1]{\begin{center}{ \it #1} \end{center}}

\newcommand{\pubblock}{\rightline{
			\usebox{\sboxpubnumber}}}
\newenvironment{Abstract}{\begin{quotation}  }{\end{quotation}}
\newenvironment{Presented}{\begin{quotation} \begin{center}
             PRESENTED AT\end{center}\bigskip
      \begin{center}\begin{large}}{\end{large}\end{center}
      \end{quotation}}
\newcommand{\Acknowledgements}{\bigskip  \bigskip \begin{center} \begin{large}
             \bf ACKNOWLEDGEMENTS \end{large}\end{center}}

\begin{document}

\begin{titlepage}
\pubdate{\today}                    
\pubnumber{} 

\vfill
\Title{Dark Matter Detection Rates In 
SUGRA Models}
\vfill
\Author{R. Arnowitt and B. Dutta}
\Address{Center For Theoretical Physics, Department of Physics, \\
Texas A$\&$M University, College Station TX 77843-4242}
\vfill
\vfill
\vfill
\begin{Abstract}Direct detection of Milky Way wimps are discussed within the framework of 
R-parity conserving SUGRA models with grand unification at $M_G$. Two 
questions are discussed: what SUGRA models can account for the DAMA data if 
this data is confirmed, and is the full SUGRA parameter space accessible to 
future planned detectors. Experimental constraints of the Higgs mass bound, 
the $b\rightarrow s\gamma$ bound, relic density constraints (including all 
co-annihilation channels), etc. are imposed. In addition, the effect of the 
possible muon $g - 2$ anomaly are examined. For mSUGRA, we find that the 
Higgs mass and $b\rightarrow s\gamma$ constraint puts a lower bound 
$m_{1/2} >$ (300 - 
400) GeV (i.e. $m_{\tilde\chi^0_1} >$ (120 - 160) GeV) for $\tan\beta < $50, and thus the largest 
theoretical neutralino-proton cross sections still lie significantly below 
the DAMA  3$\sigma$ lower bound. (Predictions for $\tan\beta> 50$ become 
sensitive to the precise value of $m_t$ and $m_b$.) If in addition one imposes 
the muon anomaly constraint, $\mu$ must be positive and an upper bound of 
$m_{1/2}<$ 850 GeV for $\tan\beta < 50$ is obtained. More generally, if $\mu >0$ and 
$m_{1/2} < 1$ TeV, the cross sections are $\stackrel{>}{\sim} 10^{-10}$ pb, and hence this parameter space 
would be mostly accessible to planned high sensitivity detectors. For 
non-minimal SUGRA models, the cross sections can be considerably larger, 
and a simple SU(5) model with non-universal soft breaking in the Higgs and 
third generation is seen to give cross sections in the DAMA range for 
$\tan\beta \stackrel{>}{\sim}$ 15 with $m_{\tilde\chi^0_1} > 80$ GeV, and minimum cross sections 
$\stackrel{>}{\sim}10^{-10}$ pb 
for $\mu >$ 0. 
\end{Abstract}
\vfill
\begin{Presented}
    COSMO-01 \\
    Rovaniemi, Finland, \\
    August 29 -- September 4, 2001
\end{Presented}
\vfill
\end{titlepage}
\def\thefootnote{\fnsymbol{footnote}}
\setcounter{footnote}{0}

\section{Introduction}It is generally agreed that a large part of the Milky Way galaxy is made of 
dark matter, i.e. current estimates of the dark matter energy density are 
$\rho_{\rm DM}\simeq (0.3 - 0.5){\rm GeV/cm^3}$. However, what this dark matter is made of 
is much less clear, and many suggestions exist. We will here assume a SUSY 
explanation, and in particular assume that the dark matter particle is the 
lightest supersymmetric particle (the LSP),  the lightest neutralino, 
$\tilde\chi^0_1$. A number of ways have been suggested to detect SUSY wimps: One may 
look for annihilation in the halo of the Galaxy leading to
 $\tilde\chi^0_1+ \tilde\chi^0_1\rightarrow 
e^+,\, \bar p + X$ and search for the anti-particle signals. Alternately, one 
expects neutralinos to be captured by the Sun or Earth, fall into their 
center and there annihilate. Muon neutrinos could then escape, and one 
could look for the signal $\tilde\chi^0_1+ \tilde\chi^0_1\rightarrow 
\nu_{\mu} + X$ with neutrino telescopes 
on Earth. Finally, there is the possibility of direct detection by the 
scattering of incident neutralinos by terrestrial nuclear targets.

The theoretical expectations for these possible signals depends upon the 
model chosen. The most general SUSY model, the MSSM, contains over 100 free 
parameters (63 real parameters!), and so it does not have great predictive 
power. The supergravity (SUGRA) models with grand unification at the GUT 
scale $M_G \simeq2\times 10^{16}$ GeV \cite{sugra1,sugra2}, apply to a wide range of phomenena, and 
are relatively predictive. We will consider here then the SUGRA models, 
both those with universal soft breaking (mSUGRA) and non-universal soft 
breaking at the GUT scale. Our models will also possess R-parity 
invariance, to guarantee the existence of a stable LSP dark matter 
candidate. In particular we examine the gravity mediated models. (Other 
possibilities such as gauge mediated soft breaking have difficulty in 
constructing a viable dark matter candidate, while the anomaly mediated 
models appear to be in contradiction with the recent Brookhaven muon 
magnetic moment data.)

Concerning possible signals, there has been  indications of an excess of 
$\bar p$ and $e^+$ events in the halo, indicating possible halo $\tilde\chi^0_1$ annihilations. 
This data is quite interesting and has led to some recent analyses 
\cite{kane,gondolo}, 
though it is difficult to be sure of all the astrophysical backgrounds. 
Analyses have also been given of the possibilities for present or future 
neutrino telescopes (AMANDA, Ice Cube, ANTARES) to see energetic $\nu_{\mu}$ from 
the Sun \cite{barger, bottino}. Thus \cite{barger} finds that Ice Cube and ANTARES can be sensitive 
to $\tilde\chi^0_1$- annihilation for non-universal SUGRA models but only for large 
$\tan\beta$ i.e. $\tan\beta \stackrel{>}{\sim}35$. The most promising way of detecting halo 
wimps for a wide range of parameters remains, then, direct detection by 
terrestrial targets, and we will restrict the discussion to this 
possibility from now on.

Experiments for direct detection of Milky Way wimps is now entering a new 
(and exciting) phase. There exists the DAMA results with data that 
indicates the observation of the annual modulation signal (at the 4-$\sigma $
level), and the CDMS exclusion curves which appear to exclude much of the 
allowed DAMA region. However, there are a number of new experiments that
should be able to clarify matters in the relatively near future. Thus DAMA 
is constructing an upgrade to a 250 kg NaI detector (from their current 95 
kg detector). Two other detectors that will be completed in the near 
future, GENIUS-TF and ZEPLIN II, should be able to give independent 
observations of whether the annual modulation effect exists, and when CDMS 
is moved to the SOUDAN mine, its exclusion curves will be sensitive to the 
full DAMA allowed region. Further, in the more distant future, a number of 
new detectors are being planned with much higher sensitivity, e.g. GENIUS, 
Cryoarray, ZEPLIN IV, and CUORE.

In view of this, we consider here two questions: (1) If the DAMA data is 
confirmed, what SUGRA models can account for the relatively large 
neutralino-proton cross sections this data implies, and (2) what are the 
lowest SUGRA cross sections theoretically predicted, and will the planned 
detectors be able to scan the full SUGRA parameter space. We will see that 
the DAMA data, if confirmed, would exclude the mSUGRA model (but be consistent
with non-universal models), and if the 
Brookhaven $g_{\mu}- 2$ anomaly is valid, future detectors will be able to test 
SUGRA models over the full parameter space and hence be competitive with 
accelerator experiments.
\section{Experimental and Theoretical Constraints}

One of the important features of SUGRA models is that they apply to a wide 
range of phenomena, and data from these already restrict the parameter 
space and thus strengthen the predictions. We list here the main 
experimental constraints that we use.\\
(1) Higgs mass\\
The LEP data \cite{higgs} places the lower bound of $m_h  > 114$ GeV. However, 
theoretical calculation of the light Higgs mass, $m_h$, still have an 
uncertainty of about 3 GeV. Thus we will conservatively interpret the 
experimental bound to require that the theoretically calculated mass obey 
$m_h > 111$ GeV.\\
(2) $b \rightarrow s  \gamma$ branching ratio\\
  The CLEO data \cite{bsgamma} has both systematic and theoretical uncertainties. We 
therefore take a relative broad range around the CLEO central value of
 \begin{equation} 1.8 \times 10^{-4} \leq B(B \rightarrow X_s\gamma) \leq 4.5 \times 
10^{-4}\end{equation}\\                                                                  
(3) $\tilde\chi^0_1$ relic density

The cosmic microwave background radiation (CMB) and other data now give a 
relatively constrained value for the cosmic mass density of dark matter. 
This is measured in terms of the value of $\Omega_{\rm DM} h^2$.  Here 
$\Omega_{\rm DM} = 
\rho/ \rho_c$  where $\rho$ is mass density of dark matter, and $\rho_c = 
3H^2/8\pi G_N$ is the critical density to close the universe ($H = 
(100km/sMpc)h$ is the current Hubble constant, $G_N$ is the Newton constant). 
A recent analysis gives $\Omega_{\rm DM} h^2 = 0.139\pm0.026$ \cite{turner}. 
We take a 2.5$\sigma$ 
range around the mean:\begin{equation}
0.07\leq\Omega_{\rm DM} h^2\leq0.021\end{equation}
In looking for the minimum cross section, we conservatively extend this 
range to be from 0.025 to 0.25.\\
(4) Recently, the BNL E821 experiment \cite{BNL} has observed a 2.6$\sigma$ anomaly 
in the muon magnetic moment $a_\mu = (g_\mu -2)/2$. Such an
anomaly was predicted some time ago for mSUGRA models \cite{g-2,g-22} and we will 
thus interpret the affect as being due to SUGRA. We take here a 2$\sigma$ 
range around the E821 central value\begin{equation}
11 \times 10^{-10}\leq a_\mu^{\rm SUGRA}\leq 75\times 10^{-10}
\end{equation}
While this effect is not yet certain (and there  is some debate as to the 
size of the Standard Model hadronic contribution, see e.g.\cite{melnikov}), it has a 
significant impact on dark matter prediction. We will therefore include it 
in our analysis,  but show explicitly what effects it produces.  Further 
data, currently being analyzed by E821, should reduce both the experimental 
error and theoretical uncertainty by more than a factor of two, and results 
should be available early next year \footnote{Recently, two papers have
appeared (Knecht et.al. (hep-ph/0111059), Hayakawa and Kinoshita (hep-ph/0112102)) implying a sign error in the scattering of light by light
contribution to $a_{\mu}$. This reduces the 2.6$\sigma$ effect to 1.6$\sigma$,
our lower bound then becomes 1$\sigma$ below the central value. We note that
with the further data, the E821 experiment would be expected to see an $\simeq
4\sigma$ effect, if the reduced central value were to remain unchanged}.

In order to carry out accurate calculations, it is necessary to include a 
number of theoretical corrections, and we list some of these here.
(1) Two loop gauge and one loop Yukawa renormalization group equations 
(RGE) are used from $M_G$ to the electroweak scale $M_{\rm EW} $ and QCD RGE are used 
below for the light quark contributions. We chose $M_{\rm EW} = 
(m_{\tilde t_L} m_{\tilde t_R})^{1/2}$.
(2) Two loop and pole mass corrections are included in the calculation of $m_h$.
(3)  One loop corrections to $m_b$ and $m_\tau$ \cite{rattazi,carena} are included which are 
important for large $\tan\beta$.
(4) Large $\tan\beta$ NLO SUSY corrections to $b\rightarrow s 
\gamma$ \cite{degrassi,carena2} are included.
(5) All stau-neutralino co-annihilation channels are included in the relic 
density calculation, this analysis being done for both small and large 
$\tan\beta$ \cite{bdutta,ellis,gomez}.
We note that we do not include Yukawa unification or proton decay 
constraints, as these depend sensitively on post-GUT physics about which 
little is known. (For example, string theory with e. g. an SU(5) GUT group 
requires grand unification of the gauge coupling constants, just as SUGRA 
theory does, but the Yukawa or proton decay constraints of SUGRA need not 
apply \cite{green}.)

\section{ mSUGRA Model}

The minimal supergravity model, mSUGRA, is the most predictive of the 
models as it depends on only four additional parameters and one sign. It 
is convenient to chose these parameters as follows: $m_0$, the scalar soft 
breaking mass at $M_G$; $m_{1/2}$, the gaugino soft breaking mass as $M_G$. (We note 
here the approximate formulae that 
$m_{\tilde\chi^0_1} \stackrel{\sim}{=} 0.4 m_{1/2}$, the chargino 
mass  $m_{\tilde\chi^{\pm}_1} \stackrel{\sim}{=} 0.8 m_{1/2}$, and the gluino mass 
$m_{\tilde g}\stackrel{\sim}{=} 2.5 m_{1/2}$.); $A_0$, the 
cubic soft breaking mass at $M_G$; $\tan\beta = <H_2>/<H_1>$ at the electroweak 
scale (where $H_{(2,1)}$ gives rise to up,down quark masses); and the sign of $\mu$, 
the Higgs mixing parameter (which appears in the superpotential as $\mu H_1 
H_2$). We examine  these parameters over the following range: $m_0$, 
$m_{1/2}\leq 1$ TeV,  $2 \leq\tan\beta \leq50$, and $A_0\leq 4 m_{1/2}$. The range of $m_0$ and 
$m_1/2$ is sufficient to explore the full range that will be accessible to 
the LHC (and we will see that if the $g_\mu -2$ anomaly is real, it will cover 
the full parameter space). There is a small additional allowed region  of very high 
$\tan\beta$  i. e. $\tan\beta \stackrel{\sim}{=}50 - 55$. However, as we will see below that 
predictions in this region are very sensitive to the precise values of the t 
and b quark masses which are not at present known.

In general, the neutralino-nucleus scattering cross section has a spin 
dependent part and a spin independent part. However, the transition from quark
to proton scattering for heavy nuclei, the 
spin independent cross section dominates, and this allows one to extract 
(to a good approximation)  $\sigma_{\tilde\chi^0_1-p}$, the neutralino-proton cross section, 
from any dark matter detector data. The basic quark diagrams for 
neutralino-quark scattering, $\sigma_{\tilde\chi^0_1-q}$, are s-channel scattering through 
first generation squarks, and t-channel scattering through h and H (where H 
is the heavy neutral Higgs boson). One can determine $\sigma_{\tilde\chi^0_1-p}$ then from 
the basic quark scattering. However, this can involve errors 
perhaps of 
a factor of $\stackrel{<}{\sim} $ 2  from uncertainties in the strange quark content of the proton and the 
pion-nucleon $\sigma$ term, $\sigma_{\pi-N}$. (We use here the most recent 
evaluations of the $\sigma$ term, $\sigma_{\pi-N} = 65$ MeV 
\cite{olsson,pavan}. Earlier 
evaluations, $\sigma_{\pi-N}\stackrel{\sim}{=}45$ MeV,would reduce the cross sections by perhaps 
a factor of 2.)
\begin{figure}[htb]
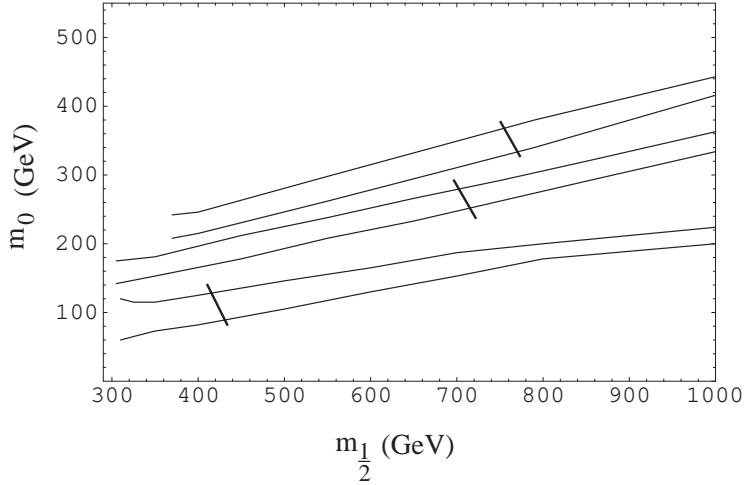

 \centerline{ \DESepsf(edmt16.epsf width 10 cm) }
\caption {\label{fig1} Corridors in the $m_0 - m_{1/2}$ plane allowed by the relic density 
constraints for (bottom to top) $\tan\beta = 10$, 30, 40, $A_0 = 0$ and $\mu > 0$. 
The lower bound on $ m_{1/2}$ is due to the $m_h$ lower bound for 
$\tan\beta$ = 10, due 
to the $b\rightarrow s \gamma$ bound for $\tan\beta =$ 40, while both these contribute 
equally for $\tan\beta = 30$. The short lines cutting the channels represent 
upper bound from the $g_\mu - 2$ experiment. [24]}
\end{figure}
\begin{figure}[htb]
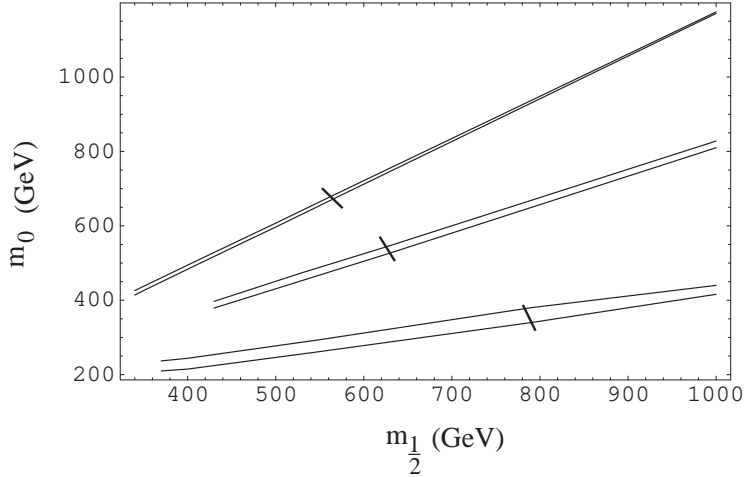

\centerline{ \DESepsf(adhs3.epsf  width 10 cm) }
\caption {\label{fig2}
 Corridors in the $m_0 - m_{1/2}$ plane allowed by the relic density 
constraint for $\tan\beta = 40,\, \mu > 0$ and (bottom to top) 
$A_0 = 0,\, -2 m_{1/2}$, 
$4m_{1/2}$. the curves terminate at the lower end due to the 
$b \rightarrow s\gamma$ 
constraint except for$ A_0 = 4 m_{1/2}$ which terminates due to the $m_h$ 
constraint. The short lines cutting the corridors represent the upper bound 
on $m_{1/2}$ due to the $g_\mu -2$ experiment. [25]}
\end{figure}
 
The parameter space for calculations of the above detector cross section 
are constrained significantly by the requirement that the theory predict 
the right amount of relic neutralino dark matter. In the early universe, 
two neutralinos can annihilate to Standard Model particles through 
s-channel  $Z$, $h$, $H$ and $A$ poles ($A$ is the CP odd Higgs boson), and through 
t-channel sfermion diagrams. However, if a second particle becomes nearly 
degenerate with the $\tilde\chi^0_1$, one must include it in the early universe 
annihilation processes, leading to the phenomena of co-annihilation. In 
SUGRA models with universal gaugino masses at the GUT scale, this 
``accidental" near degeneracy occurs naturally for the light stau, $\tilde\tau_1$. To 
see this semi-quantitatively, we note that for low and intermediate 
$\tan\beta$, the RGE can be solved for the right selectron, $\tilde e_R$, and 
$\tilde\chi^0_1$ masses 
to give at the electroweak scale:
\begin{eqnarray}
 m_{\tilde e_R}&=&m_0^2 + 0.15 m_{1/2}^2  - sin^2\theta_W M_W^2
 \cos2\beta\\\nonumber m_{\tilde\chi^0_1}&=& 0.16m_{1/2}^2 
\end{eqnarray}
where the numerical factors come from solving the RGE. The last term is 
about (35GeV)$^2$, and so for $m_0 =0$, the $\tilde e_R$ becomes degenerate with the 
$\tilde\chi^0$ 
at $m_{1/2}\stackrel{\sim}{=} 350$ GeV, and thus co-annihilation begins at 
$m_{1/2} \stackrel{\sim}{=}(350 - 
400)$GeV. As $m_{1/2}$ increases, $m_0$ must be raised in lock step (to keep 
$m_{\tilde e_R} > 
m_{\tilde\chi^0_1}$). More, precisely, it is the light stau, which is the lightest 
slepton, that dominates the co-annihilation phenomena. However, one ends up 
with corridors in the $m_0 - m_{1/2}$ plane with allowed relic density, with $m_0$ 
closely correlated with $m_{1/2}$ and $m_0$ increasing as $m_{1/2}$ increases.

We now turn to the other experimental constraints. The Higgs mass bound and 
the $b\rightarrow s \gamma$ bounds produce comparable constraints on the parameter 
space, the Higgs mass generally being dominant for low $\tan\beta$ and the 
$b\rightarrow s\gamma$ for high $\tan\beta$. Over the entire $\tan\beta$ range these constraints 
produce a lower bound on $m_{1/2}$:
\begin{equation}
                            m_{1/2}\geq (300 - 400) {\rm GeV}
\end{equation}
corresponding to a lower bound on the neutralino mass of
\begin{equation}
                                  m_{\tilde\chi^0_1}  \geq (120-160)  {\rm GeV}
\end{equation}

This means that most of the parameter space is in the 
$\tilde\tau-\tilde\chi^0_1$ 
co-annihilation domain in the relic density calculation. Thus relic density 
bounds then imply $m_0$ is approximately determined by $m_{1/2}$. 
\begin{figure}[htb]
\centerline{ \DESepsf(cosmotalk2.epsf  width 10 cm) }
\caption {\label{fig3}
Allowed region in the $m_0 - m_{1/2}$ plane for $\tan\beta =$ 50, $A_0 = 0$,
 $\mu > 
0$ for $m_t = 175$ GeV, $m_b = 4.25$. The lower and upper bounds on $m_{1/2}$ are due 
to the $b\rightarrow s\gamma$ and $a _\mu$ constraints respectively.}
\end{figure}

\begin{figure}[htb]
\centerline{ \DESepsf(adscosmo5.epsf  width 10 cm) }
\caption {\label{fig4}Allowed regions in the $m_0 - m_{1/2}$ plane for $\tan\beta = 50$, $A_0 =$ 0, 
$\mu > 0$ for $m_t = 170$ GeV, $m_b = 4.25$ GeV. The lower and upper bounds  on 
$m_{1/2}$ 
are due to the $b \rightarrow s \gamma$ and $a _\mu$ constraints respectively.}
\end{figure}
If we now include the Brookhaven $g_\mu - 2$ constraint we obtain additional 
restrictions on the parameter space. First the sign of $a_\mu^{\rm SUGRA}$ implies 
that $\mu > 0$. Further, the lower bound on $a_\mu^{\rm SUGRA}$ 
implies an upper bound 
on $m_{1/2}$. Thus for $A_0 = 0$, one has $m_{1/2} < 440$ GeV for $\tan\beta = 10$, 
$m_{1/2} < 
790$ for $\tan\beta = 40$ and $m_{1/2} < 850$ for $\tan\beta = 50$. Thus assuming the 
Brookhaven E821 data, we obtain both an upper bound on $m_{1/2}$ and a lower 
bound on $m_{1/2}$. The reduction of the parameter space for small $\tan\beta$
then puts a lower bound on $\tan\beta$. We find 
\begin{equation}
\tan\beta>7(5),\,\,{\rm for} \,\,A_0=0(-4 m_{1/2})
\end{equation}

Some of the above results are illustrated in Figs. 1 and 2. Fig. 1 shows 
the allowed regions (which are the co-annihilation channels) in the $m_0 - 
m_{1/2}$ plane for $\tan\beta = 10$, 30, 40 (from bottom to top) for $A_0 = 0$ 
\cite{bdutta1}. 
The lower $m_{1/2}$ bounds are due to the $m_h$ bound for $\tan\beta =$ 10, and due to 
the $b\rightarrow s  \gamma$ bound for $\tan\beta = 40$. (They are equally constraining 
for $\tan\beta =$ 30.) The $g_\mu - 2$ upper bounds are given by the short lines 
cutting the channels. One sees that the co-annihilation corridors occur for 
higher $m_0$ as $\tan\beta$ increases. Fig. 2 shows the $A_0$ dependence for the case 
of $\tan\beta = 40$. One sees that the co-annihilation corridors lie higher in 
$m_0$ for $A_0$ different from zero.

For $\tan\beta > 50$, a new phenomena arises due to the fact that the heavy 
Higgs (H, A) become light, and can contribute to the annihilation cross 
section. Results in this domain become very sensitive to the precise values 
of $m_t$ and $m_b$ \cite{ellis2}. We illustrate this in Figs. 3 and 4. Fig. 3 gives the 
allowed region in the $m_0 - m_{1/2}$ plane for $\tan\beta = 50$, $A_0 = 0$ using the 
central experimental values of $m_t = 175$ GeV and $m_b = 4.25$. However, 
if $m_t$ 
is reduced by 1$\sigma$ to 170 GeV, one sees from Fig. 4, that the shape of the 
allowed region is significantly modified, and a new corridor is opened at 
low $m_{1/2}$ rising to $m_0 > 1$ TeV.

We see that the parameter space has become highly restricted, both at the 
lower end and the upper end, and this should help in clarifying the 
predictions for dark matter detection. The general features of 
$\sigma_{\tilde\chi^0_1-p}$
that explain its properties are that $\sigma_{\tilde\chi^0_1-p}$ increases with increasing 
$\tan\beta$, and decreases with increasing $m_{1/2}$ and  $m_0$ (and 
as we've seen,$m_{1/2}$ and  $m_0$ move together). Thus the maximum values of $\sigma_{\tilde\chi^0_1-p}$
 are generally expected 
to occur for high $\tan\beta$, and low $m_{1/2}$, $m_0$. We now return to our two 
questions of Sec. 1: in mSUGRA, how large can cross sections get, i. e. can 
they accommodate the DAMA data, and how small can they become, i. e. will 
future detectors be able to scan the entire parameter space. The current 
DAMA annual modulation signal is in the range \cite{belli}
\begin{equation}
(\sigma_{\rm wimp})_{\rm DAMA}  \stackrel{\sim}{=}  ( 10^{-5} - 10^{-6}) {\rm
pb}\,\, 
{\rm and}\,\,  
m_{\rm wimp}\leq (100 - 300) {\rm GeV}
\end{equation}
where the uncertainty in the upper bound on the wimp mass being due in part
to
astronomical uncertainties about the Milky Way. We compare this now  to the 
mSUGRA predictions. Fig. 5 shows the $A_0$ dependence of the cross sections 
for $\tan\beta = 40$. Note that $A_0 = 0$ gives the largest cross section. We 
see that it is the low $m_{1/2}$ bound that prevents the cross section from 
becoming large, (and in fact if the bound on $m_h$ were to increase to 120 
GeV, the lower bound on $m_{1/2}$ would rise to 200 GeV for $A_0 = -2m_{1/2}$, to 215 
GeV for $A_0 = 0$, and to 246 GeV for $A_0 = 4 m_{1/2}$ further reducing the maximum 
cross sections). Fig. 6 shows the cross sections for $\tan\beta=10$, 30, 40 and
50 
for $A_0 = 0$, $m_t = 175$ GeV. The largest cross  sections occur for the largest 
$\tan\beta$ and smallest $m_{1/2}$, as expected. We find in general,
\begin{equation}
\sigma_{\tilde\chi^0_1-p}<  (0.07)\times10^{-6} pb \,\, {\rm for}\,\,
\tan\beta  <  50    
\end{equation}
which is well below the DAMA bound. Thus it is not possible for the mSUGRA 
model to accommodate the DAMA data. We note that central in this result is 
the $m_h$ and $b \rightarrow s\gamma$ constraints which put a lower bound on 
$m_{1/2}$ and 
hence require $m_{\tilde\chi^0_1}$ to be relatively high.

\begin{figure}[htb]
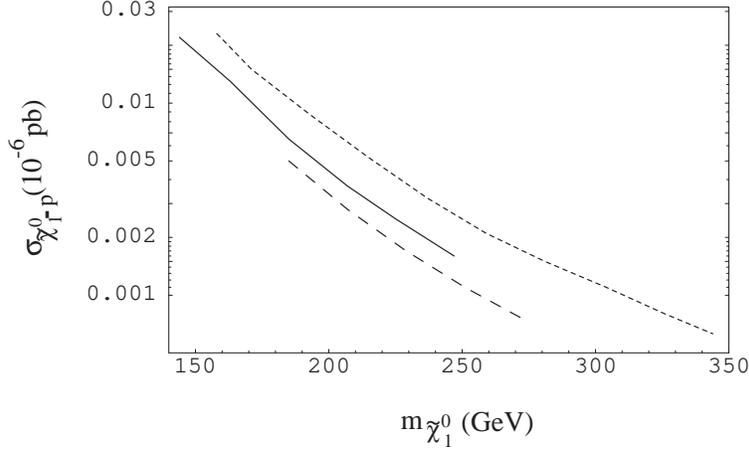

 \centerline{ \DESepsf(adhs2.epsf width 10 cm) }
\caption {\label{fig5} $\sigma_{\tilde\chi^0_1-p}$ for $\tan\beta = 40$, $\mu >0$, 
for (bottom to top), $A_0= -2m_{1/2}$, $A_0 = 4 m_{1/2}$, 
$A_0 = 0$. 
The lower bounds on $m_{\tilde\chi^0_1}$ are as in Fig. 2.}\end{figure}

\begin{figure}[htb]
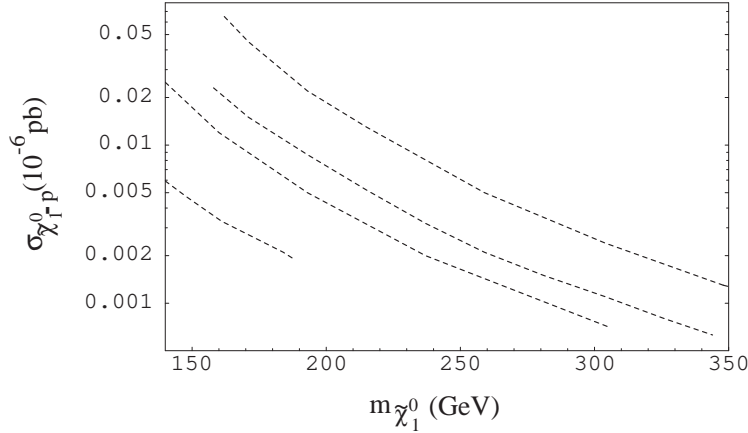

 \centerline{ \DESepsf(cosmotalk4.epsf width 10 cm) }
\caption {\label{fig6} $\sigma_{\tilde\chi^0_1-p}$ for $A_0 = 0$, $m_t = 175$ GeV, $m_b = 4.25$, 
for $\tan\beta = 10$, 
30, 40 50 (from bottom to top). The lower bound on $m_{\tilde\chi^0_1}$  is 
due to the $b \rightarrow s\gamma$ constraint for $\tan\beta=$ 30, 40, 50, 
and from the $m_h$ constraint for 
$\tan\beta =$10.}\end{figure}
\begin{figure}[htb]
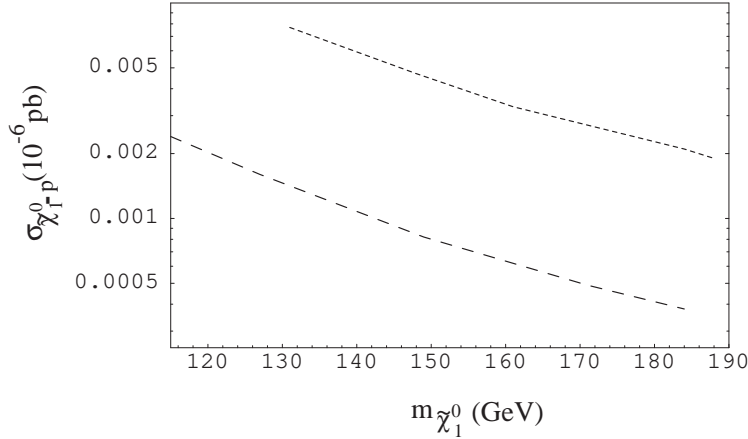

\centerline{ \DESepsf(adhs1.epsf  width 10 cm) }
\caption {\label{fig7} $\sigma_{\tilde{\chi}_1^0-p}$ for $\tan\beta = 10$, 
$A_0 = 0$ (upper curve), $A_0 = -4 m_{1/2}$ (lower curve). The
lower bound on $m_{\tilde{\chi}_1^0}$ is due to the $m_h$ bound for $A_0 = 0$, 
and the
$b \rightarrow  s\gamma$ constraint for 
$A_0 = -4 m_{1/2}$. }
\end{figure}

We turn now to the second question, what are the minimum values possible 
for $\sigma_{\tilde\chi^0_1-p}$, and will future detectors being planned be able to observe 
them. Minimum cross sections arise from low $\tan\beta$ and large $m_{1/2}$, $m_{0}$. 
Here the question of the validity of the Brookhaven $g_\mu -2$ data plays an 
important role in two ways. First this data would eliminate the $\mu < 0$ 
possibility and for $\mu < 0$, there are  accidental cancellations that can 
reduce $\sigma_{\tilde\chi^0_1-p}$
  to well below $10^{-10}$ pb over wide ranges of parameters 
\cite{bdutta,ellis3}, which would make this part of the parameter space experimentally 
unaccessible. Thus these extremely low cross sections would be eliminated. 
Second, it is the lower bound on $a_\mu^{\rm SUGRA}$ that produces the upper bound 
on $m_{1/2}$, and this effect is larger for smaller $\tan\beta$ (see e. g. Fig. 6) 
moderating the reduction of the cross section as $\tan\beta$ decreases. Fig. 7 
shows $\sigma_{\tilde\chi^0_1-p}$ for $\tan\beta = 10$, $A_0 = 0$ 
(upper curve), $-4m_{1/2}$ (lower 
curve). We see here that $\sigma_{\tilde\chi^0_1-p}> 4\times10^{-10}$ pb.
 The parameter space 
ends for $\tan\beta =$ 7 (5) for $A_0 = 0 (-4m_{1/2})$  and in general one finds 
$\sigma_{\tilde\chi^0_1-p}\geq 10^{-10}$ pb.

Future detectors hope to obtain a sensitivity of 
$\sigma_{\tilde\chi^0_1-p} \stackrel{\sim}{=}( 10^{-9} - 
10^{-10} )$ pb. Thus they should be able to scan most of the mSUGRA 
parameter space. If this sensitivity is achievable, dark matter experiments 
would be competitive with accelerator experiments.
\section{ Non-Universal Models}

We next investigate the possibility that the DAMA data gets confirmed, and 
the large neutralino -proton cross sections it implies are correct. To 
accommodate this theoretically, one can consider models with non-universal 
soft breaking at $M_G$ in the Higgs and third generation squarks and 
sleptons. A convenient parameterization is
\begin{eqnarray} 
m_{H_{1}}^{\ 2}&=&m_{0}^{2}(1+\delta_{1}); 
\quad m_{H_{2}}^{\ 2}=m_{0}^{2}(1+ \delta_{2});\nonumber \\ m_{q_{L}}^{\
2}&=&m_{0}^{2}(1+\delta_{3}); \quad m_{t_{R}}^{\ 2}=m_{0}^{2}(1+\delta_{4});
\quad m_{\tau_{R}}^{\ 2}=m_{0}^{2}(1+\delta_{5});  \nonumber \\ m_{b_{R}}^{\
2}&=&m_{0}^{2}(1+\delta_{6}); \quad m_{l_{L}}^{\ 2}=m_{0}^{2}(1+\delta_{7}).
\label{eq26}
\end{eqnarray}
with $-1 \leq \delta_i\leq 1$. While this appears to involve introducing a large 
number of new parameters, one can understand the effects they produce 
relatively simply, as much of the physics of what occurs is governed by the 
$\mu$ parameter. One can qualitatively see the effects of the 
non-universalities for low and intermediate $\tan\beta$ where one may solve the
RGE for $\mu^2$ analytically \cite{nath}:
\begin{eqnarray}
\mu^2&=&{t^2\over{t^2-1}}\left[({{1-3 D_0}\over 2}+{1\over
t^2})+{{1-D_0}\over2}(\delta_3+\delta_4)\right. \nonumber \\ 
&-&\left.{{1+D_0}\over2}\delta_2+{\delta_1\over
t^2}\right]m_0^2+{\rm {universal\,parts\,+\,loop \, corrections}}. \label{eq28}
\end{eqnarray}
where $t = \tan\beta$ and $D_0 = 1 - (m_t/200 \sin2\beta)^2$. One see that 
$D_0$ is 
small (i. e. $D_0\stackrel{\sim}{=} 0.25$) and that the universal contribution to the $m_0^2$ term 
nearly cancels out. Thus one does not need a large amount of 
non-universality for the $\delta_i$ contributions  to dominate. Now the important point is 
that if $\mu^2$ decreases (increasing the Higgsino content of $\tilde\chi^0_1$) 
$\sigma_{\tilde\chi^0_1-p}$ 
increases, and similarly if $\mu^2$ increases, $\sigma_{\tilde\chi^0_1-p}$
 decreases (though by 
not as large an amount). From Eq(11), then we see  that the cross section 
increases if $\delta_{1,3,4} <0, \delta_2 >0$. We consider here a simple $SU(5)$ 
model where $\delta_{10}\equiv \delta_3 = \delta_4 = \delta _5$, and 
$\delta_{\bar 5}\equiv  
\delta _6 = \delta _7$. We chose $\delta_1 = -1$, $\delta_2 = 1$, with 
$\delta_{\bar 5} >0$ 
and $\delta_{10} < 0$. With these choices, $\mu^2$ is reduced and hence the cross 
section is increased. One must, however, make sure the relic density 
constraint is simultaneously satisfied. This occurs due to the fact that
the $\delta_i$ chosen also lowers $m_A$. Thus the RGE for $m_A$ give for low and 
intermediate $\tan\beta$:
\begin{eqnarray}
m_A^2&=&{{t^2+1}\over{t^2-1}}\left[{{3(1- D_0)}\over
2}+{{1-D_0}\over2}(\delta_3+\delta_4)\right. \nonumber  \\
&-&\left.{{1+D_0}\over2}\delta_2+\delta_1\right]m_0^2+{\rm
{universal\,parts\,+\,loop \, corrections}}.   \label{eq29}
\end{eqnarray}

Thus $m_A$ is simultaneously lowered opening an annihilation channel for  
${\tilde\chi^0_1}$ 
through an $A$ s-channel pole, and allowing the relic density constraint to 
be satisfied. What is happening is similar to what happens in mSUGRA at 
$\tan\beta\geq 50$, but for the non-universal model this effect can happen at 
low $\tan\beta$ and with a significantly larger effect. Fig. 8 shows the 
maximum value of $\sigma_{\tilde\chi^0_1-p}$ for the model discussed above for 
$\tan\beta = 12$ 
(lower curve), 18 (upper curve).  One sees that one can generate 
$\sigma_{\tilde\chi^0_1-p}$ 
in the DAMA domain for $\tan\beta \stackrel{>}{\sim}15$. Note also that the lower bound on 
$m_{\tilde\chi^0_1}$
is greatly reduced, since the Higgs mass and $b\rightarrow s\gamma$ constraints are 
much less confining for the non-universal model. (Again a lower wimp mass 
would be in better accord with the DAMA results.) Thus if the DAMA data is 
confirmed by other groups, this could point to the validity of 
non-universal SUGRA models.

The minimal cross sections arise from low $\tan\beta$, large $m_{1/2}$, $m_0$ and 
also  from reversing the signs of the $\delta_i$. One finds for $\mu > 0$ (as 
implied by the $g_\mu - 2$ data) that $\sigma_{\tilde\chi^0_1-p}\geq 10^{-10}$pb, which as 
before is within the reach of future planned detectors.

\begin{figure}[htb]
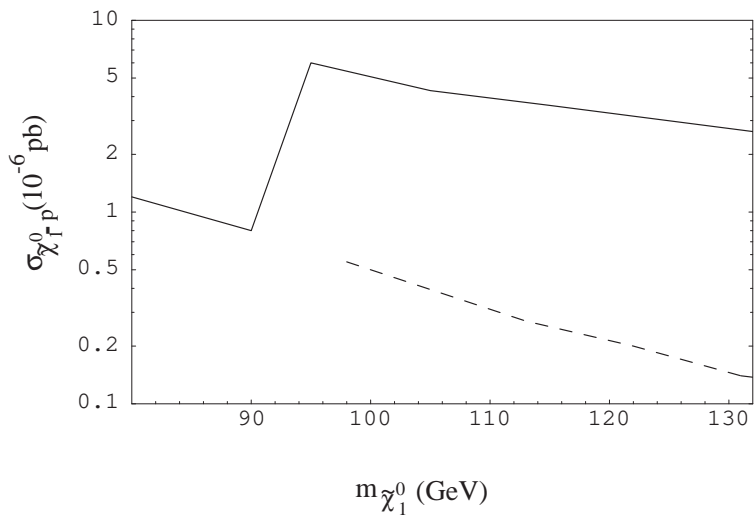

 \centerline{ \DESepsf(cosmotalk1.epsf width 10 cm) }
\caption{\label{fig8}Maximum values of $\sigma_{\tilde\chi^0_1-p}$ for $\tan\beta = $12 (lower curve), 
18(upper curve) for $\delta_2 = 1 = -\delta_1$, $\delta_{10} <0$, 
$\delta_{\bar 5} > 0$, 
$A_0 = 0$, $ \mu > 0$.}\end{figure}

\section{Conclusions}

We have discussed here the direct detection of neutralino dark matter in 
the Milky Way within the framework of R-parity conserving gravity mediated 
supergravity models with grand unification at $M_G \stackrel{\sim}{=}
2 \times10^{16}$ GeV. In 
particular, we examined two questions: if the DAMA data is confirmed, can 
SUGRA models accommodate the large cross sections implied by the data, and 
also, what are the lowest cross section the SUGRA theory predicts and will 
future planned detectors be able to scan the entire parameter space.

There are now many experimental constraints that allow one to make clearer 
predictions for these models. Thus for mSUGRA, the Higgs mass and 
$b\rightarrow s\gamma$ constraints create a lower bound of $m_{1/2} \geq (300-400)$ GeV (i.e. 
$m_{\tilde\chi^0_1}\geq(120- 160)$GeV ), which puts the parameter space mainly in the 
co-annihilation domain, and (approximately) determines $m_0$ in terms of 
$m_{1/2}$ 
(for fixed $\tan\beta$ and $A_0$). We find then that the largest cross section is 
significantly below the lower bound of the DAMA data for $\tan\beta <$ 
50.  (For $\tan\beta \geq$ 50, predictions are sensitive to the precise values of 
$m_t$ and $m_b$.) If the muon $g - 2$ anomaly is confirmed by future data, then 
significant further reduction of the parameter space occurs. One has that 
$\mu$ is positive, eliminating the possibility of very small cross sections. 
In addition, this effect produces an upper bound of $m_{1/2} < 850$ GeV for 
$\tan\beta <$ 50 as well as a lower bound on $\tan\beta$ of $\tan\beta > $7 (5) for 
$A_0$ 
$= 0 (-4m_{1/2})$. We find then over the remaining parameter space that cross 
sections are greater than $\simeq 10^{-10}$ pb and hence most of the parameter 
space will be accessible to the future planned experiments. This result 
also holds without the $g_\mu - 2$ anomaly for the $\mu >0$, $ m_{1/2}\leq 1$ TeV part of 
the parameter space.

For non-minimal SUGRA models, with non-universal soft breaking in the Higgs 
and third generation squark and slepton sectors, one can generate models 
with significantly larger neutralino-proton cross sections. Thus for a 
simple SU(5) model, one finds it is possible to find cross sections in the 
DAMA region for $\tan\beta > 15$, and also the lower bound on the neutralino 
mass can be reduced to 80 GeV. These models satisfy the relic density 
constraint by simultaneously reducing the $A$ mass (even for the relatively 
low $\tan\beta$ above), and one also finds that for the $\mu >0$ part of the 
parameter space the cross sections should not fall significantly below 
$10^{-10}$ pb, and thus be experimentally accessible.
\Acknowledgements This work was supported in part by National Science Foundation Grant 
PHY-0101015.


\begin{thebibliography}{99}


\bibitem{sugra1}A.H. Chamseddine, R. Arnowitt, P. Nath,
{\PRL} {\bf 49} {(1982)} {970}.
\bibitem{sugra2} R. Barbieri, S. Ferrara, C.A. Savoy,
{\PLB}{\bf 119} {(1982)} {343}; L. Hall, J. Lykken, S. Weinberg,
{\PRD}{\bf 27} {(1983)}  {(2359)}; P. Nath, R. Arnowitt, A.H. Chamseddine,
{\NPB}{\bf 227} {(1983)}   {(121)}.
\bibitem{kane} G.L. Kane, L.-T. Wang, J.D. Wells, hep-ph/0108138.
\bibitem{gondolo} E.A. Baltz, J.Edsjo, K. Freese,  P.Gondolo,
astro-ph/0109318.
\bibitem{barger}V.Barger, F. Halzen, D. Hooper, C. Kao, hep-ph/0105182. 
\bibitem{bottino} A. Bottino, N. Fornengo, S. Scopel, F. Donato,
hep-ph/0105233.

\bibitem{higgs} P.~Igo-Kemenes, LEPC meeting, November 3, 2000 
(http://lephiggs.web.cern. ch/LEPHIGGS/talks/index.html). 

\bibitem{bsgamma} M. Alam etal {\PRL}{\bf 74}  {(1995)} {(2885)}.


\bibitem{turner}M. Turner, hep-ph/0106035.

\bibitem{BNL}H.N. Brown et.al., Muon (g-2) Collaboration,
\PRL {\bf 86}, (2001) 2227. 
\bibitem{g-2} T.C. Yuan, R. Arnowitt, A.H. Chamseddine, P. Nath,
{\ZPC} {\bf 26} {(1984)} {407}.

\bibitem{g-22} D.A. Kosower, L.M. Krauss, N. Sakai,
{\PLB} {\bf 133} {(1983)} {305}.

\bibitem{melnikov} K. Melnikov, {\IJMPA} {\bf 16} {(2001)} {4591}.

\bibitem{rattazi}R. Rattazi, U. Sarid, {\PRD}{\bf 53}  {(1996)} {(1553)}.
 
\bibitem{carena}M. Carena, M. Olechowski, S. Pokorski, C. Wagner
{\NPB}{\bf 426} {(1994)} {(269)}.

\bibitem{degrassi}G. Degrassi, P. Gambino, G. Giudice,
{JHEP}{\bf 0012} {(2000)} {009}.
\bibitem{carena2}M. Carena, D. Garcia, U. Nierste, C. Wagner,
{\PLB} {\bf 499} {(2001)} {141}.

\bibitem{bdutta} R. Arnowitt, B. Dutta, Y. Santoso, hep-ph/0010244; hep-ph/0101020;
{\NPB} {\bf 606}  {(2001)} {59}.

\bibitem{ellis} J Ellis, T. Falk, G. Ganis, K. Olive, M. Srednicki,
{\PLB} {\bf 570} {(2001)} {236};  
J. Ellis, T. Falk, K. Olive, {\PLB} {\bf 444} {(367)} {1998};
J. Ellis, T. Falk, K. Olive, M. Srednicki, {\AP} {\bf 13} {(2000)} {181};
Erratum-ibid.{\bf 15}, (2001) 413.
\bibitem{gomez} M. Gomez, J. Vergados, hep-ph/0012020; 
M. Gomez, G. Lazarides, C.
Pallis, {\PRD}{\bf 61} {(2000)} {123512}; Phys. Lett. {\bf B487}, (2000) {313};
 L. Roszkowski, R. Austri, T. Nihei, JHEP {\bf 0108},  {(2001)}  024; 
 A. Lahanas, D. Nanopoulos, V. Spanos,  {\PLB} {\bf 518} {(2001)} 
 {94} and talk by V. Spanos at this conference.
\bibitem{green} See e.g., M.B. Green, J.H. Schwarz, E. Witten, ``Superstring
theory", vol.2, Sec.16.4 (Cambridge University Press, Cambridge, 1987).

\bibitem{olsson}M. Olsson, {\PLB} {\bf 482} {(2000)} {50}. 
\bibitem{pavan}M. Pavan, R. Arndt, I.
Strakovsky, R. Workman, PiN Newslett. {\bf 15} (1999) 118.
\bibitem{bdutta1}R. Arnowitt, B. Dutta,   Y. Santoso,
{\PRD} {\bf 64} {(2001)} {113010}.
\bibitem{bdutta2}R. Arnowitt, B. Dutta, B. Hu,  Y. Santoso,
Phys. Lett. {\bf B505},  (2001) 177.
\bibitem{ellis2}J. Ellis, T. Falk, G. Ganis, T. Vergata, K. Olive,
M. Srednicki, \PLB {\bf 510}, (2001) 236; J. Ellis, S. Heinemeyer, 
K. Olive,
G. Weiglein, \PLB {\bf 515}, (2001), 348. 
 
\bibitem{belli}P. Belli et.al. hep-ph/0112018,  talk at this conference.

\bibitem{ellis3}J. Ellis, A. Ferstl, K. A. Olive, 
{\PLB} {\bf 481} {(2001)} {304}; {\PRD} {\bf 63} {(2001)} {065016}.

\bibitem{nath}R. Arnowitt, P. Nath, {\PRD} {\bf 56} {(1997)} {2820}.
\end{thebibliography}
\end{document}